\documentclass[preprint]{aastex}
 
\begin{document}
  
\title{A First Look at White Dwarf - M Dwarf Pairs in the Sloan Digital Sky
Survey (SDSS)}
%\footnote{Based on 
%observations obtained with the Sloan Digital Sky Survey and with the
% Apache Point Observatory (APO) 3.5m telescope, which are owned and operated
%by the Astrophysical Research Consortium (ARC)}}

\received{9/15/2002}
\accepted{2/13/2003} 

\author{Sean N. Raymond\altaffilmark{1,2}, Paula Szkody\altaffilmark{2}, Suzanne
L. Hawley\altaffilmark{2}, Scott F. Anderson\altaffilmark{2},
J. Brinkman\altaffilmark{3}, Kevin R. Covey\altaffilmark{2},
P. M. McGehee\altaffilmark{4}, D. P. Schneider\altaffilmark{5},
Andrew A. West\altaffilmark{2}, D. G. York\altaffilmark{6}}

\altaffiltext{1}{Corresponding Author: raymond@astro.washington.edu}
\altaffiltext{2}{Department of Astronomy, University of Washington, Box 351580,
Seattle, WA 98195}
\altaffiltext{3}{Apache Point Observatory, P.O. Box 59, Sunspot, NM 88349-0059}
\altaffiltext{4}{Los Alamos National Laboratory, Los Alamos, NM 87545}
\altaffiltext{5}{Dept. of Astronomy and Astrophysics, The Pennsylvania State
University, University Park, PA 16802}
\altaffiltext{6}{Dept. of Astronomy and Astrophysics and Enrico Fermi
Institute, 5640 S. Ellis Ave, Chicago, IL 60637}

\index{White Dwarf / M Dwarf pairs}
\index{binaries!interacting}
\index{SDSS}
\index{low mass stars}
\index{Magnetic Activity}

\begin{abstract}

We have identified 109
White Dwarf (WD) - M dwarf pairs in the Sloan Digital Sky Survey (SDSS) with
{\it g}
$<$ 20th magnitude.  For each system we determined the temperature of the WD
primary and the spectral type of the M dwarf secondary.  Using H$\alpha$ emission as
a proxy for the chromospheric activity level of the M dwarf, we investigated
correlations between the activity level and properties of the system.
Compared with field M dwarfs (Hawley et al. 1996), we see a slightly higher
active fraction of early-type stars, with activity levels similar to the field.

We have conducted followup observations at the ARC 3.5m telescope to
obtain radial velocity information and to search for short period binaries,
which may be on the verge of interacting.
We report on one system with a 4.1 hour period, and several additional systems
with significant velocity variations.

\end{abstract}

\noindent keywords: binaries: spectroscopic  --- Sloan Digital Sky Survey
(SDSS) --- stars: activity --- 
\indent stars: late-type --- white dwarfs

\section{Introduction}

The formation and evolution of a low mass star in a binary system is a
common phenomenon, and can lead to interesting results such as
cataclysmic variables (Warner 1995), which are the most numerous UV and X-ray sources
in the Galaxy, and Type Ia supernovae (Langer et al 2000), which are standard candles
for cosmology.  While the low mass secondaries play a pivotal role in
determining the evolution of the system, it is not well understood how the
binary environment affects the evolution and properties of the secondary.

Studies of pre-cataclysmic variable stars (or Pre-Cataclysmic Binaries --
PCBs) (Bond et al 1985; Hillwig et al 2000)
envision the following scenario for their evolution.  The
more massive primary in a wide binary star system evolves through the
giant and planetary nebula phases, leading to significant mass loss and a
``common envelope'' encompassing both primary and secondary.   A large amount
of the system's orbital angular momentum is lost during this phase, leading to
a much closer binary orbit.  Tidal interactions in the binary system 
will have led to synchronous rotation of the secondary for separations closer
than 5-7 R$_{g}$, where R$_{g}$ is the radius of the primary during its giant
phase (Soker 2002), corresponding to an orbital period of $\sim$50 days
(assuming an M5 secondary).  The remaining binary may undergo further orbital
decay via magnetic braking or gravitational radiation (see Warner 1995).
Eventually, systems with very short
periods will interact and become cataclysmic variables (CVs).  CVs therefore
represent only the short-period subset of all these binaries.  In this paper
we add to the growing database of PCBs, focusing on faint WD/M pairs.  
We hope to eventually build a large sample of PCBs which represents
various
phases of  binary evolution, in order to test the validity of the above
scenario, and the efficiency of the various physical mechanisms for angular
momentum loss.  

Note that we define the WD, the original high-mass component, as the
primary in these systems.  This contrasts with previous surveys, many of
which defined the primary as the brighter star in a given filter, whether it
be the WD or the M dwarf.  

The Sloan Digital Sky Survey (SDSS; York et al 2000) database, with its
accurate photometry in five broad bands covering the entire optical spectrum,
is an outstanding resource for finding these pre-interacting systems.  We have
identified 109 WD/M binaries with g $<$ 20, using data available through
11/14/2001.  Our efficiency at finding these objects using our
current photometric targeting algorithm is $\sim$ 60 \%, and our sample is
augmented by serendipitous observations obtained through other target
selection pipelines  (e.g. quasars).  Forty three of our systems are contained in the
June 2001 SDSS Early Data Release (Stoughton et al 2002).  

Currently known are 42 PCBs (Hillwig et al 2000), and roughly 800 widely
separated, non-interacting WD/M binaries (Silvestri, Oswalt \& Hawley 2002 and
references therein).  We expect to
accumulate a sample of 700-1000 systems by the end of the survey, including
$\sim$ 50 PCBs -- doubling the current samples of these objects.

For each of our systems, we analyzed the SDSS discovery spectrum to determine
the temperature and distance of the white dwarf primary and the spectral type
and magnetic activity level of the low mass secondary.  We are particularly
interested in studying the dependence of the magnetic activity level on
rotation rate, which increases in close binary systems where the
secondary is in synchronous rotation.  The magnetic activity of the secondary
has been widely cited as a primary driver for angular momentum loss as stars
evolve toward the period gap (Patterson 1984).  We also investigated the
influence of irradiation from the hot white dwarf on the activity level of the
secondary. 

To further compare activity levels across
spectral type and rotation rate, we need to determine which pairs are
in close binaries (thus exhibiting synchronous rotation). Population
models (de Kool 1992; Warner 1995 and references therein) indicate that about 10\% of the
WD/M pairs should be close enough to enter the common envelope stage, with periods $\le$ 4 hrs.  We targeted the most active systems for followup observations at
the 3.5m ARC telescope at Apache Point Observatory (APO), to search for radial
velocity variations, and obtain orbital periods.  For a 2 M$_{\odot}$ WD
progenitor, a secondary closer than $\sim$0.25 AU (P $\leq$ 50 days) will rotate synchronously
(Soker 1996).  The radial velocity of an M5 dwarf (M $\approx$ 0.2
M$_{\odot}$) in such a binary is $\sim$40 km~s$^{-1}$.  With our radial
velocity errors of $\pm$10 km~s$^{-1}$, we can therefore detect both
synchronously and non-synchronously rotating binaries.

In Section 2 we describe our technique for selecting WD/M
pairs from the SDSS photometric database. Section 3 presents our methods for
deriving system
parameters from the SDSS spectra.  In Section 4 we discuss the results of this
analysis, including our ongoing followup at APO.  Section 5 summarizes our findings.

\section{SDSS and APO Data}

\subsection{Photometric Selection of WD/M Pairs with SDSS}

Our goal is to find the
evolutionary precursors of CVs, systems on the verge of interacting,
or only weakly interacting.  For these systems we expect no significant
flux from an accretion disk, and the composite spectrum of the candidate pair
will be simply the superposition of the spectra of each component.
Therefore we wish to isolate objects with blue spectral energy
distributions (SEDs) at short
optical wavelengths (hot WDs), and red SEDs at long wavelengths (M dwarfs).
To accomplish this, we use the Sloan filter system (Fukugita et al 1996)
to devise a photometric targeting algorithm.

Through a trial and error procedure during the commissioning year of the SDSS,
we optimized a set of color selection criteria in the 4 SDSS colors: u - g $<$
0.45 (blue), g - r $<$ 0.7 (blue), r - i $>$ 0.3 (red), and i - z $>$ 0.4
(red).  We found that a practical magnitude cutoff of g $<$ 20 was
required to obtain targeted SDSS spectra of our candidates with a sufficient
signal to noise ratio for subsequent analysis.

Figure 1 shows the color distribution of our full sample of 109 systems.
About half of the sample was found by visual inspection of SDSS spectra
targeted for other purposes (e.g. galaxies, quasars), and many of these
objects lie outside of our color cuts.  However, we are unable to broaden the
color selection criteria further, as too many interlopers from the stellar
locus would be targeted, lowering our efficiency.  

When our color criteria are applied to all SDSS objects for which spectra have
been taken, $\sim$ 60\% of these are WD/M pairs.  We have repeat observations
of 16 systems, one of which was observed 3 times.  Note that our sample is not
at all complete, as we take an average of one spectrum per SDSS field, which
is 7 square degrees.  The completeness of and biases in our sample will be addressed in
an upcoming paper based on a larger SDSS WD/M sample.

Table 1 lists the SDSS photometry for the full sample, and Figure 2
illustrates the wide variety of systems we observed.  Panels a) and b) are
classic WD/M pairs, with SEDs that are blue
at short wavelengths and red at long wavelengths.  In some cases, the
binary spectrum is dominated by the hot WD primary (panel c) or the cool M
dwarf secondary (panels d and f), while panel e) shows nearly equal
contributions over the whole spectrum.  Panels a) and c) show very
active systems with strong H$\alpha$ emission and emission cores inside the
pressure-broadened WD Balmer absorption lines.  The WD primaries in the
systems from panels d) and f) are similar, however, d) is active while f) is
not. 

\subsection{APO Data}

We have observed 31 of our 109 pairs with the ARC 3.5-m telescope at APO, using
the high resolution grating on the Double Imaging Spectrograph (DIS), which produces spectra
in both the blue and red portions of the optical spectrum simultaneously.  The
dispersion is about 1.6\AA\ pixel$^{-1}$ in the blue, and 1.1 \AA\
pixel$^{-1}$ in the red. 

Table 3 shows information for the 12
systems with APO spectra on at least two different nights.  We tracked 5
objects for several hours, to look for variations on $\le$ 4 hour timescales.

\section{Analysis of SDSS Data}

The temperature of the WD primary in each spectrum was determined by fitting the spectrum to hydrogen WD
models (Hubeny \& Lanz, 1995).  Only one object, SDSS144258.48$+$001031.5,
showed helium absorption lines.  We fit the spectra between 3800 and
5000 \AA, a region with very little contamination from the M dwarf secondary,
to models with temperatures between 8,000K and 98,000K.  The models
are normalized to the spectrum at 4200\AA, which is a
fairly flat region of continuum between the strong H$\delta$ and H$\gamma$
absorption lines.  The best fit was found using a simple chi-squared
minimization technique.  The errors are, conservatively, $\pm$ 2,000K for temperatures
below 20,000K, and $\pm$ 4,000K above 20,000K.  Once the best fit temperature
was determined, the distance was found from the
normalization factor that matches the model flux to the spectrum, assuming a
WD radius of 8 $\times$ 10$^{8}$ cm (M$_{WD}$ = 0.6 M$_{\odot}$).  We then
subtracted the best-fit WD model from the binary spectrum to extract the
spectrum of the low mass secondary.  

The spectral types of the secondaries were determined from these extracted
spectra, using primarily the TiO5 spectral index defined in Reid, Hawley and
Gizis (1995).  This index measures the strong TiO band near 7050 \AA, and is
well correlated with spectral type over the range K7 - M6.  There are no
secondaries with later types in our sample, which is expected given the
extreme faintness of these very late M dwarfs (precluding their selection in
our sample).

The distances obtained from the WD analysis were checked using the
spectroscopic parallaxes of the M dwarf secondaries.  We obtained
absolute magnitudes for the M dwarfs as a function of spectral type using
relations from Hawley
et al (2002).  The distances from the two methods generally agreed within
20\%.  However, the spectroscopic parallax of late-type stars using SDSS
colors is under revision (West et al in preparation).  We therefore use the
distances derived from the WD temperature fitting.

The activity level of an M star is best measured by the ratio of its
H$\alpha$ luminosity to its bolometric luminosity, L$_{H\alpha}$/L$_{bol}$.  
The H$\alpha$ equivalent width (EW) of each star was measured interactively
using the IRAF\altaffilmark{6}  ``splot'' routine, as continuum placement is difficult in these subtracted spectra,
making automatic routines unreliable.  The
H$\alpha$ luminosities of our M dwarfs were calculated from the equivalent
widths using the average
H$\alpha$ continuum fluxes as a function of spectral type from the PMSU
sample (Hawley, Gizis \& Reid 1996).  The bolometric luminosities
were calculated from the spectral type averages given in Reid and Hawley 2000
({\it New Light on Dark Stars}).  Table 1 contains all of the WD and M dwarf
parameters determined from our analysis.  Table 2 bins the data by the
spectral type of the secondary.

\altaffiltext{6}{IRAF is distributed by the National Optical Astronomy Observatories, which are operated by the Association of Universities
for Research in Astronomy, Inc., under cooperative agreement with the National Science Foundation.}

\section{Results}

Figure 3 shows the temperature distribution of the white dwarfs in our
sample.  The average temperature is $\sim$16,000K, slightly
higher than the average for field WDs, which is $\sim$12,000K.  The highest
temperature seen is $\sim$44,000K $\pm$4,000K.  Thus, using the cooling curves
from Fontaine et al (2001) and assuming a mass of 0.6 M$_{\odot}$, the
youngest (hottest) WD has a post main-sequence age of less than 0.1 Gyr.  The
coolest model was 8,000K, implying a post main-sequence age between 1 and 2
Gyr. 

The average M spectral type of the low mass secondaries is $\sim$M4.5.  The
distribution of spectral types of the sample is shown in Figure 4.  The rise
toward spectral type
M4-5 is due to the mass function of the local neighborhood, while the decrease
toward later spectral types is due to the lower detection efficiency for
fainter objects.

The distribution of derived distances is shown in Figure 5, assuming radii of
8 $\times$ 10$^{8}$ cm for all WDs, which should be a valid assumption for at
least $\sim$ 80 \% of WDs (Bergeron et al 1992).  All but one of the systems
lie within 200 pc, indicating that we are sampling the local disk population.

Roughly 60\% of the systems are magnetically active (see Table 2).  This is a higher fraction
for these early-mid M dwarfs than that measured by the Palomar/MSU (PMSU)
Survey of nearby stars (Hawley, Gizis \& Reid 1996).  As noted in the PMSU Survey and for later types
by Gizis et al (2000), the fraction of
stars that are active increases with later spectral type, until virtually
every M7 star is active.  Figure 6 shows that a higher fraction of early-type
M dwarf components (M1 - M4) in
WD/M pairs are active than their field
counterparts.  The level of activity as measured by
 L$_{H\alpha}$/L$_{bol}$ (Figure 7) is similar for active M dwarfs both in the field and in
our binaries, but there is a possible suggestion that the M2 - M4 dwarfs in
the WD/M pairs are somewhat more active.  However, our sample is relatively small, with few points earlier than
M2.5 or later than M6, so these results must be confirmed by further studies
with a larger sample.

Surprisingly, Figure 8 shows only a very weak correlation between M dwarf
activity and WD temperature.  A stronger correlation would imply that
irradiation by the WD was enhancing the H$\alpha$ emission from the
M dwarf.  

The source of magnetic activity
in late type stars is not well understood, and information on the rotation
dependence of the activity would provide a constraint on existing dynamo
models.  Close binaries, with large radial velocity variations, should have
strong enough tidal effects to induce the M dwarf secondary to rotate synchronously with the
white dwarf primary, thereby significantly increasing its rotation rate.  If the M dwarf dynamo is
rotation dependent, the activity level should thus increase in these close
binary systems.

For the 12 objects with at least two followup APO spectra, we can calculate a lower limit for the 
radial velocity variation.  Figure 9 compares their activity
levels with V$_{min}$ (i.e. their minimum $v \sin i$).  No correlation is immediately
apparent, but again more data are needed to provide a significant test.
One object out of the five that we tracked for 3 continuous hours,
SDSS112909.50$+$663704.4, was found to have a period of $\sim$4.1 hours (see
Figure 10).  No clear periodicity was found in the four other systems.

\section{Conclusions}

We have compiled a sample of 109 WD/M binaries from the first year
of the Sloan Digital Sky Survey.  We expect the sample to contain 700-1000 systems by the end of the five year survey.  These systems, which are the
precursors of cataclysmic variable stars and type Ia supernovae, span a wide
range in WD temperature, M dwarf spectral type and chromospheric activity
level. 

The average temperature of our sample is $\sim$16,000K, considerably hotter than the average
temperature for field WDs.  The mean spectral class of the low-mass
secondaries is $\sim$M4-M5.  The fraction of active early-type M dwarfs in binaries
appears higher than the active fraction among field M dwarfs (Figure 6).  Their
magnetic activity levels may also be elevated relative to those seen in the field
M dwarfs (Figure 7).  More data are required to confirm and quantify these
suggestive results.

Since a higher orbital velocity implies a closer orbit, in which tidal
interactions spin up the M dwarf secondary and possibly increase the efficiency of a
rotation-dependent dynamo, we have begun to obtain orbital velocities for the
brightest, most active systems in our sample.  

Population models (de Kool 1992) predict that $\sim$ 10\% of WD/M binaries should be
pre-interacting with orbital periods of $\sim$ 4 hours.  We have discovered
one such system from the five which were followed for 3 or more hours.  Eleven
other systems show some evidence for velocity variations during scattered
observations.  We see no correlation between the minimum orbital velocity 
and the M dwarf activity level.  However, much more data are needed to reach any firm
conclusions.  Our sample size is continuously increasing, and further analysis
and followup observations are in progress.  We encourage others to pursue
velocities of the identified systems.  

Funding for the creation and distribution of the SDSS Archive has been provided by the Alfred P. Sloan
                Foundation, the Participating Institutions, the National Aeronautics and Space Administration, the
                National Science Foundation, the U.S. Department of Energy, the Japanese Monbukagakusho, and the
                Max Planck Society. The SDSS Web site is http://www.sdss.org/. 

                The SDSS is managed by the Astrophysical Research Consortium (ARC) for the Participating
                Institutions. The Participating Institutions are The University of Chicago, Fermilab, the Institute for
                Advanced Study, the Japan Participation Group, The Johns Hopkins University, Los Alamos National
                Laboratory, the Max-Planck-Institute for Astronomy (MPIA), the Max-Planck-Institute for
                Astrophysics (MPA), New Mexico State University, University of Pittsburgh, Princeton University, the
                United States Naval Observatory, and the University of
Washington.

\clearpage

\begin{figure}
\epsscale{1.}
\plottwo{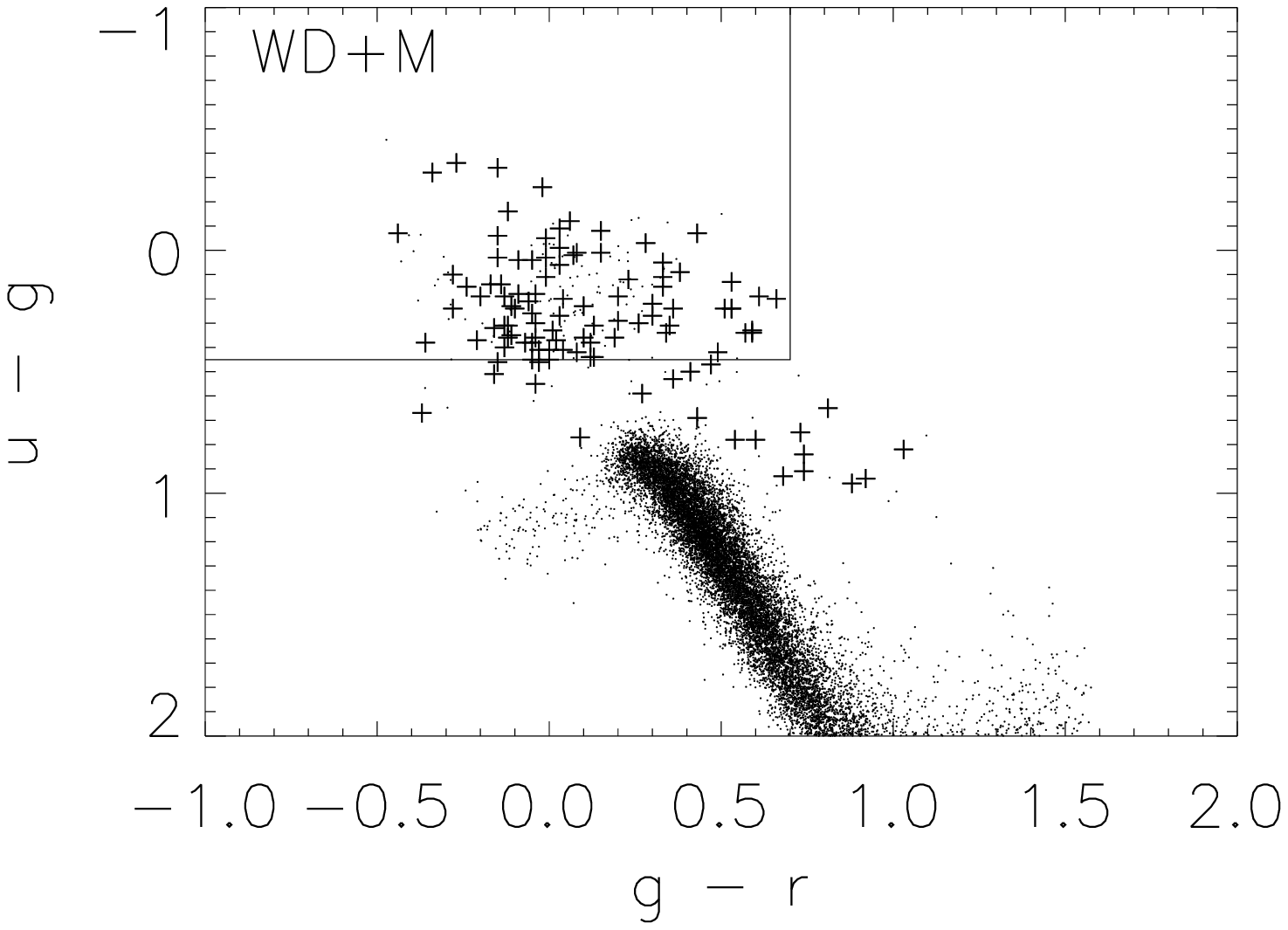}{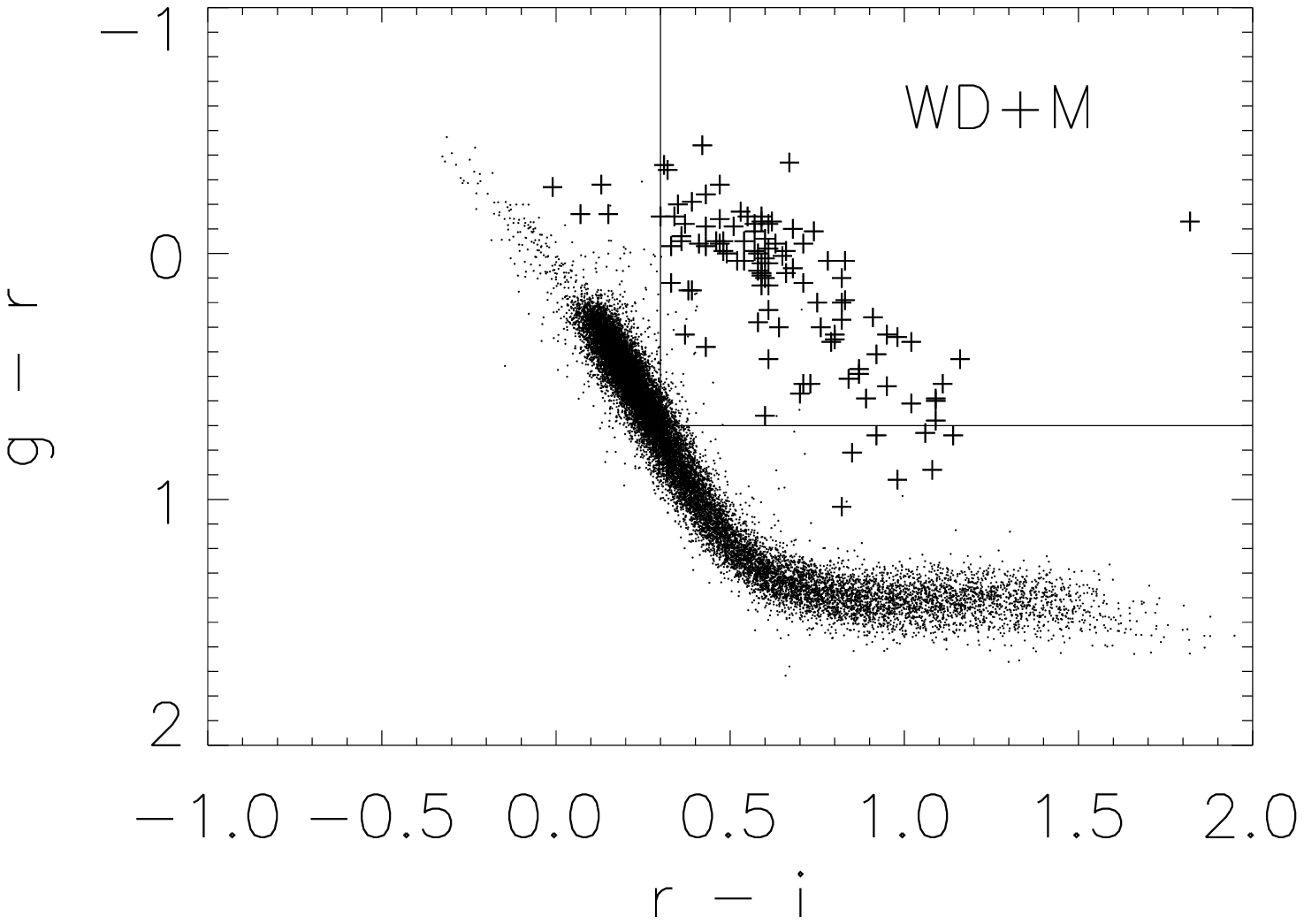}
\centerline{\plotone{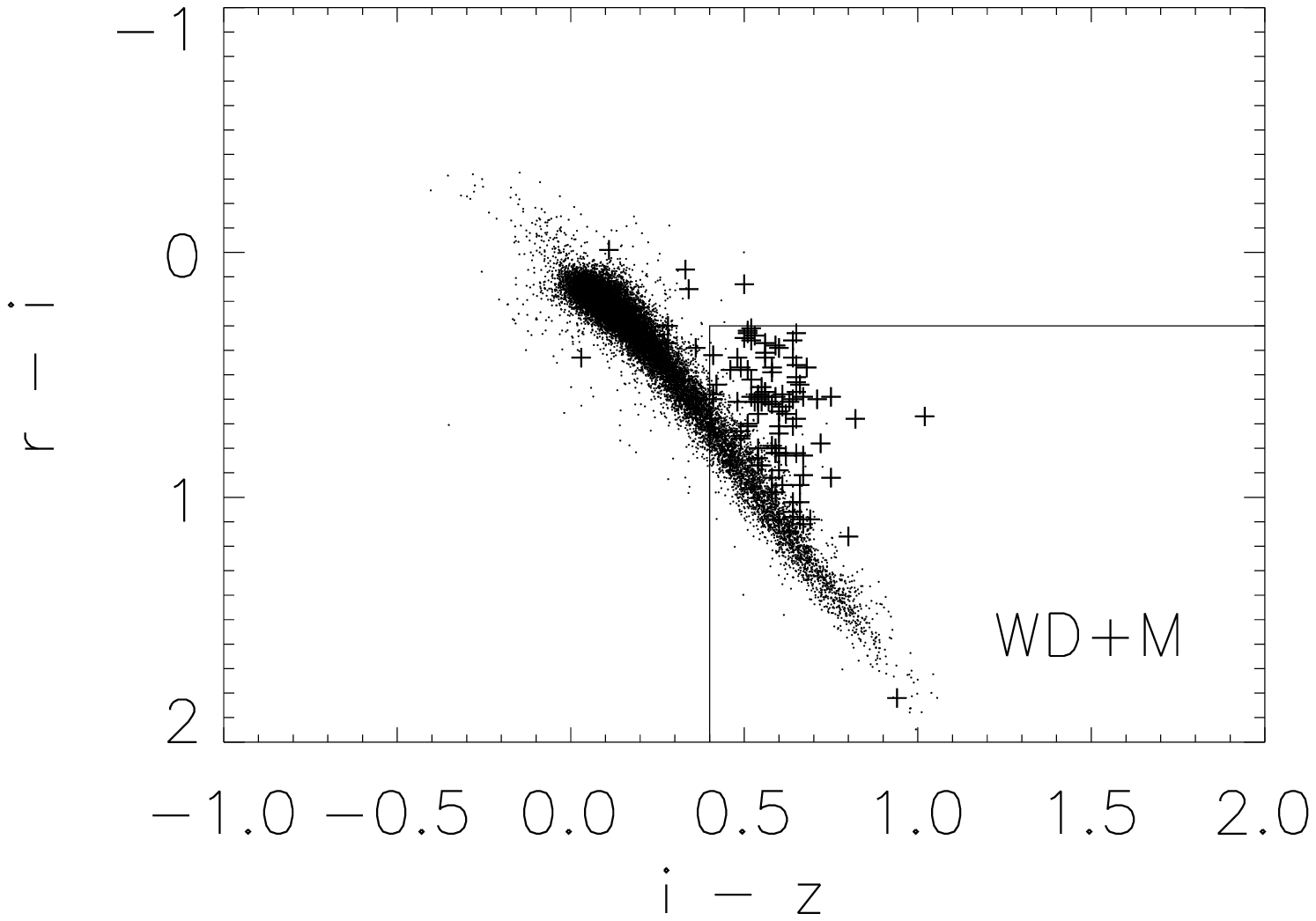}}

\caption{SDSS color-color plots of our 109 WD/M pairs.  The crosses are the
WD/M binaries while the small dots are stars defining the stellar locus.
Objects are selected which have blue colors at short wavelengths (hot WDs)
and red colors at longer wavelengths (M-dwarfs).  Our color criteria are: u -
g $<$ 0.45, g - r $<$ 0.7, r - i $>$ 0.3, and i - z $>$ 0.4.}
\end{figure}

\begin{figure}
\plotone{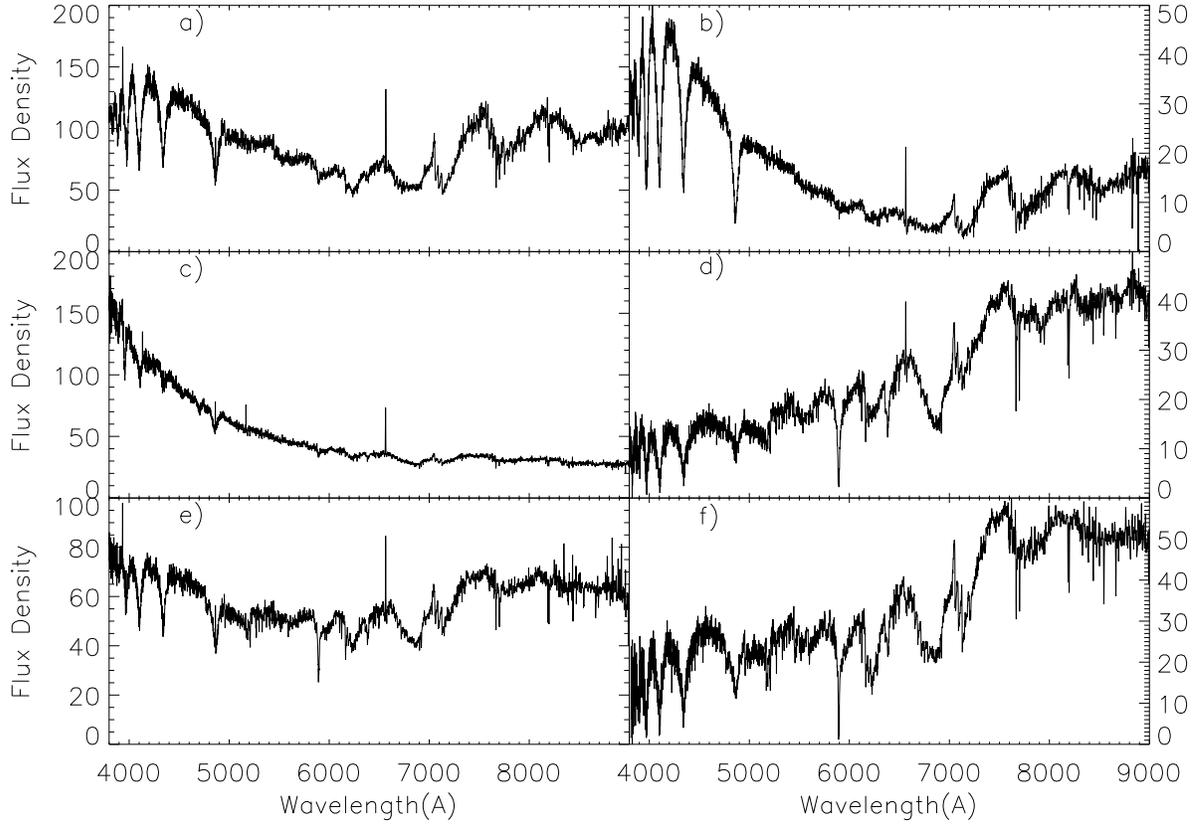}
\caption{Sample of WD/M binary spectra from our sample.  Flux density
is in units of 10$^{-17}$ ergs cm$^{-2}$ s$^{-1}$ \AA$^{-1}$.  The spectra
are: a) SDSS03067.2-003114.4, b) SDSS112909.50+663704.4, c)
SDSS001749.25-000955.4, d) SDSS172043.87+560109.3, e) SDSS231230.79+005321.7,
and f) SDSS032842.92+001749.8.  SDSS112909.50$+$663704.4 (panel b) has a period of $\sim$ 4.1 hours (see Figure 10).}
\end{figure}

\begin{figure}
\plotone{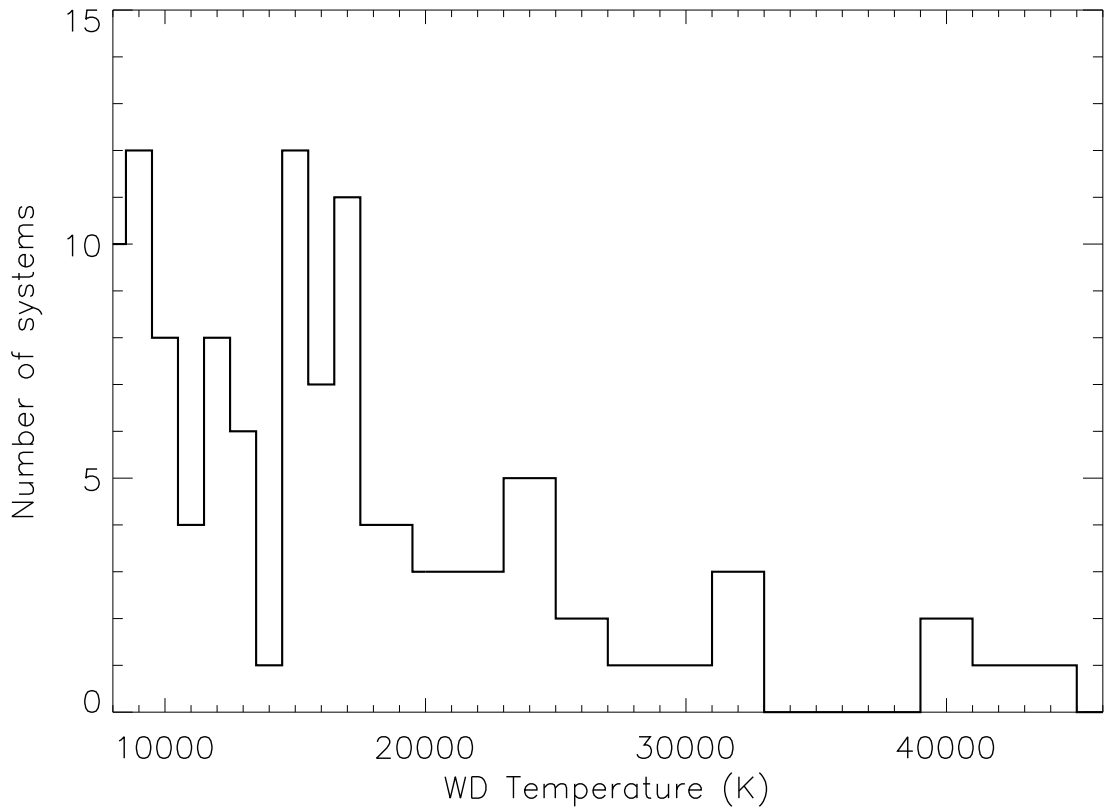}
\caption{Temperature distribution for the 109 white dwarfs in our sample.  The
average temperature is 16,000K and the hottest is 44,000K.  Our coolest WD
model had a temperature of 8,000K. }
\end{figure}

\begin{figure}
\plotone{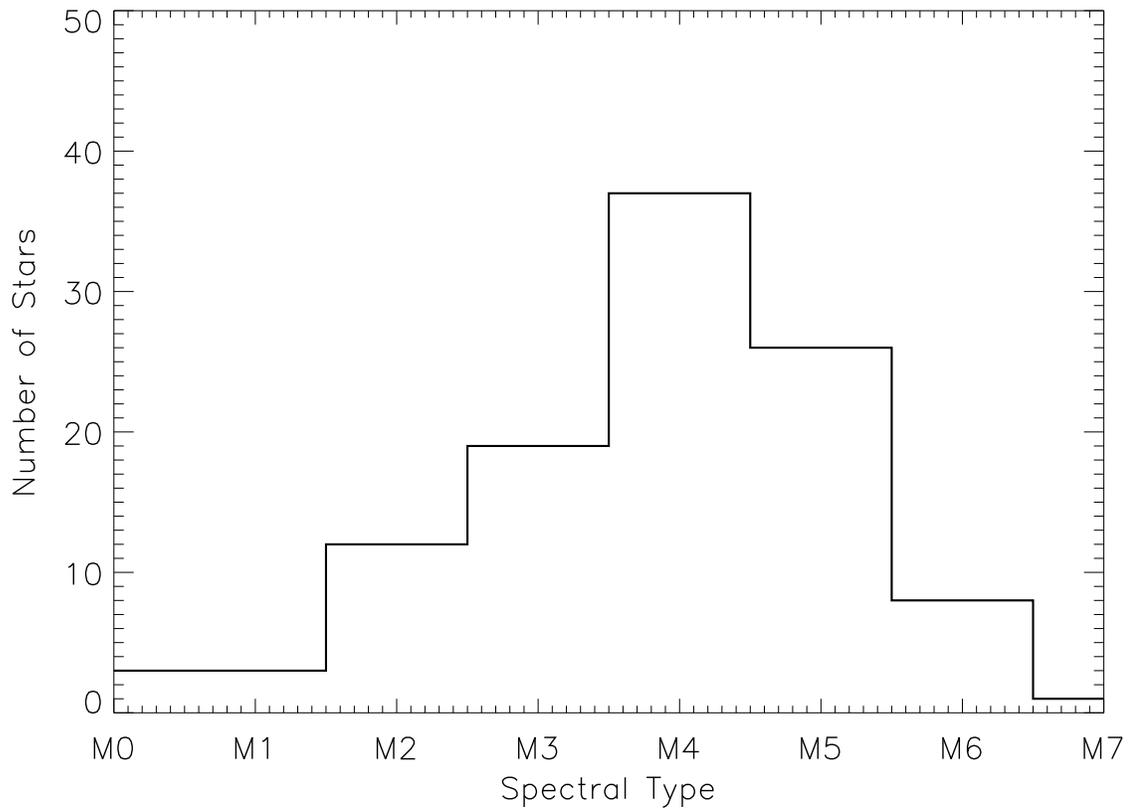}
\caption{Distribution of the M spectral types for our 109 systems, after
subtracting off the WDs.  The average spectral type is M4.3, and the errors
are roughly $\pm$ 1 spectral type.  The peak at $\sim$ M4.5 is
a result of convolving the mass function of the Galactic Disk (which increases for
later spectral types) with the detection efficiency (which
decreases for later spectral types).}
\end{figure}

\begin{figure}
\plotone{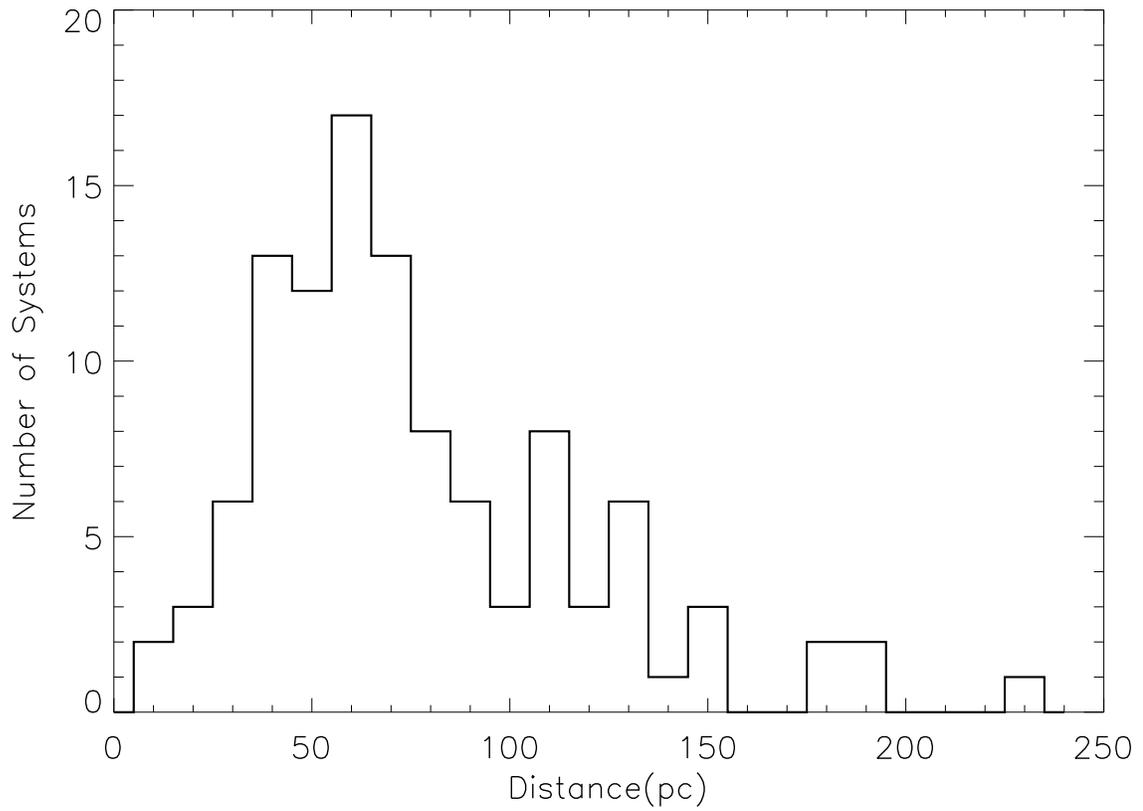}
\caption{Distribution of distances for our 109 systems.  These are calculated
from the WD model normalization factor, assuming a WD radius of 8 $\times$
10$^{8}$ cm, which is expected for a 0.6 M$_{\odot}$ WD. The average distance
is 80pc.}
\end{figure}

\begin{figure}
\plotone{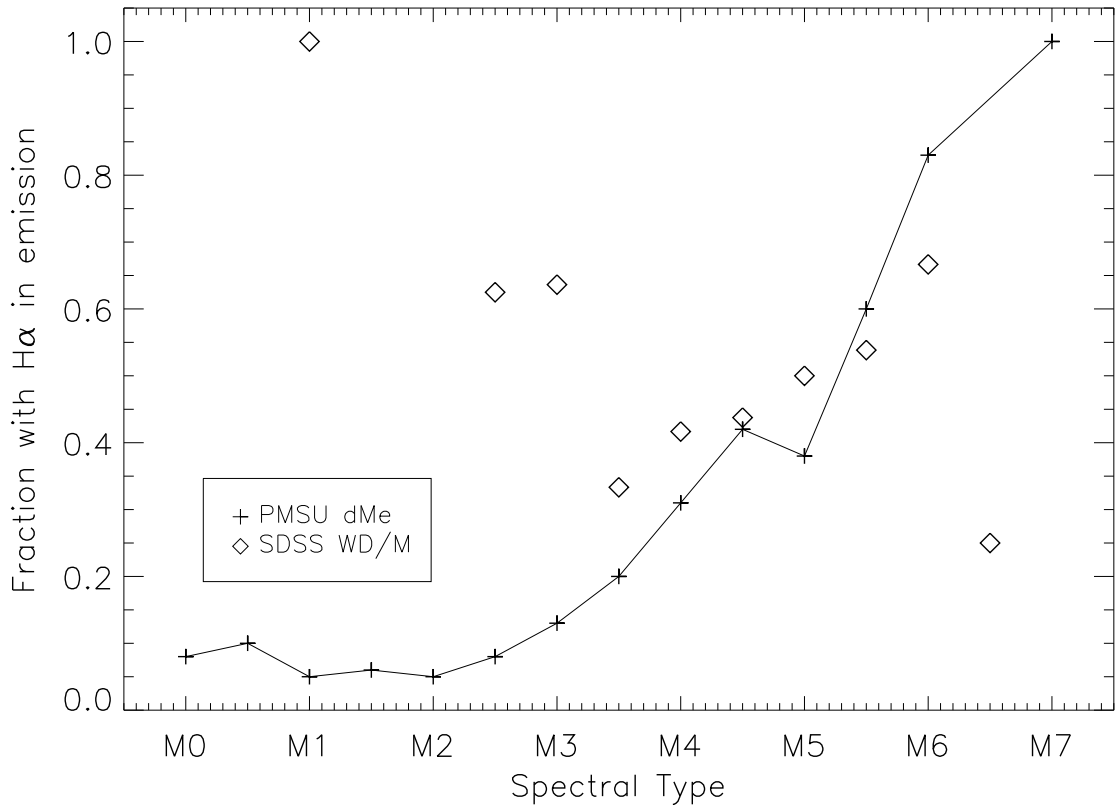}
\caption{Fraction of M dwarfs showing magnetic activity vs spectral type.  A
  higher fraction of the early-type M dwarfs (M1-M4) are active compared with
  their field counterparts.  Table 2 contains the data for Figures 6-8.}
\end{figure}

\begin{figure}
\plotone{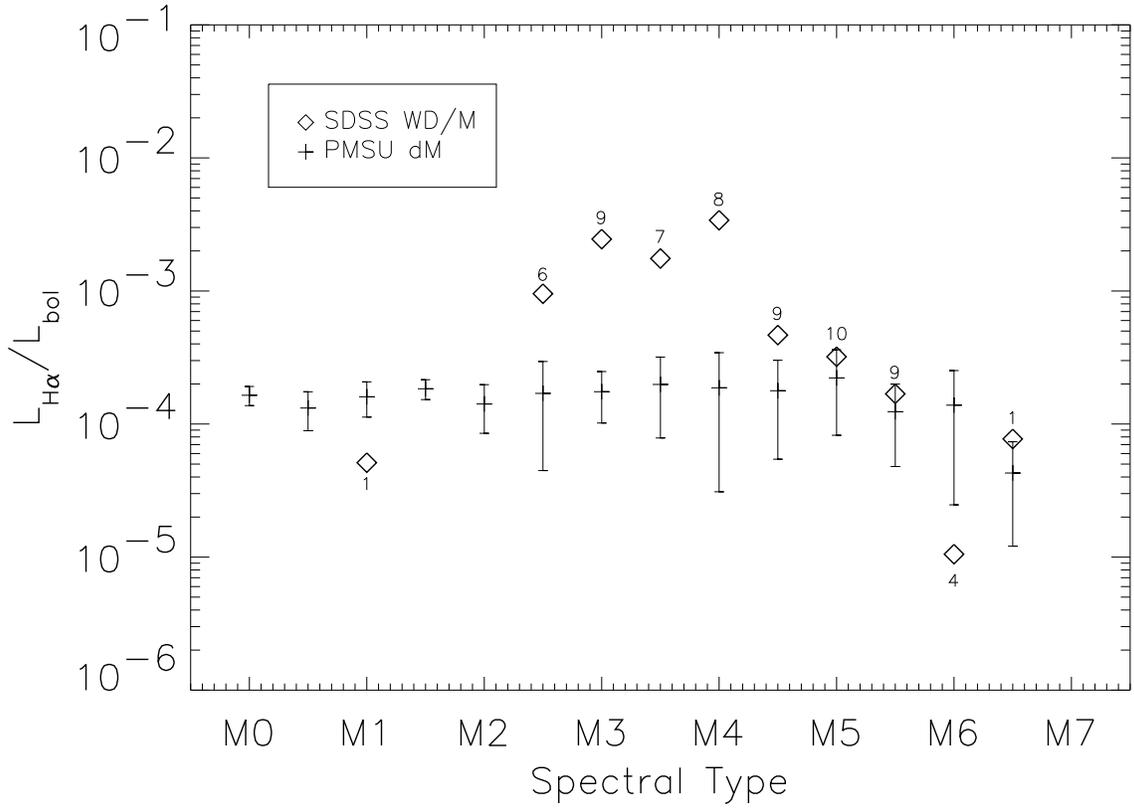}
\caption{M dwarf activity level vs spectral type for the WD/M pairs compared
with the PMSU field M dwarfs (Hawley, Gizis \& Reid 1996).  The numbers near
the diamonds indicate how many stars are in each bin. 
The uncertainties
are listed in Table 2.  It appears that the M2.5-M4 secondaries have a higher average
activity level than their field counterparts, but the number of objects is
still too small to make a strong statement.}
\end{figure}

\begin{figure}
\plotone{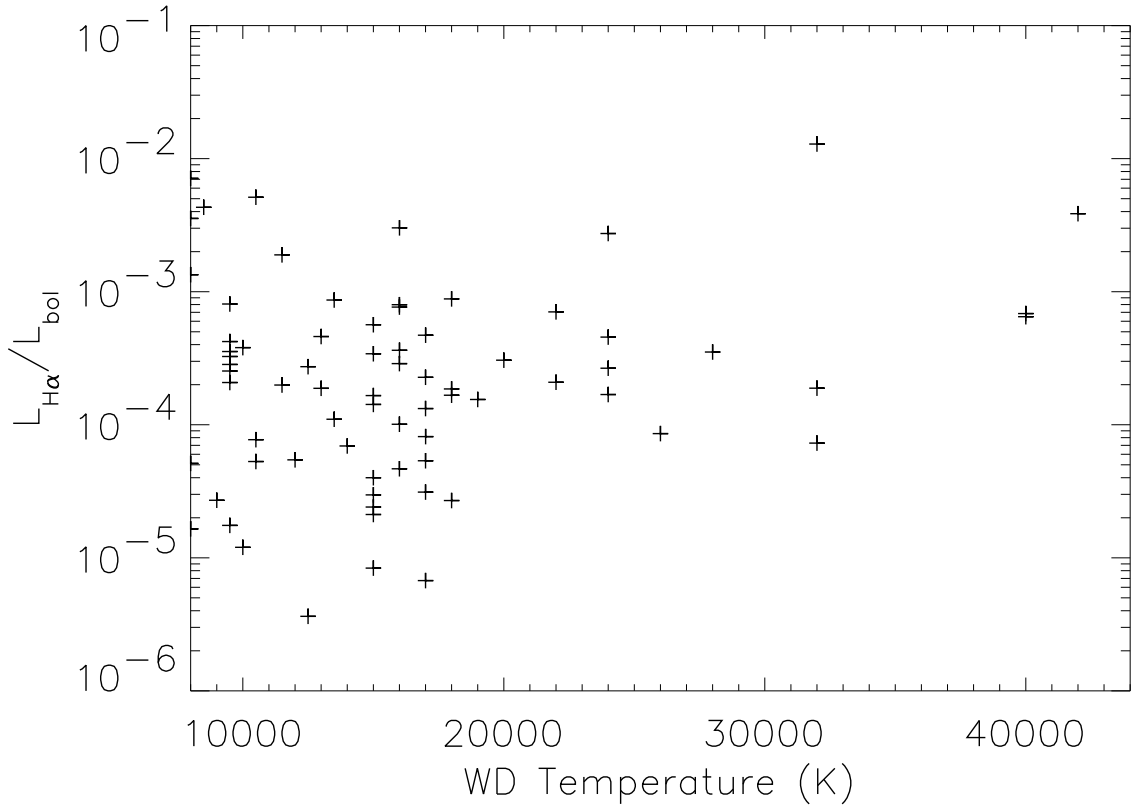}
\caption{M dwarf activity vs WD temperature.  The average error bar is
  5$\times$10$^{-4}$.  There is a weak correlation
between WD temperature and M dwarf activity level, which may be confirmed with
a larger sample.  There may also be an age-activity effect (Gizis, Reid \&
Hawley 2002), as older white dwarfs have lower temperatures.}
\end{figure}

\begin{figure}
\plotone{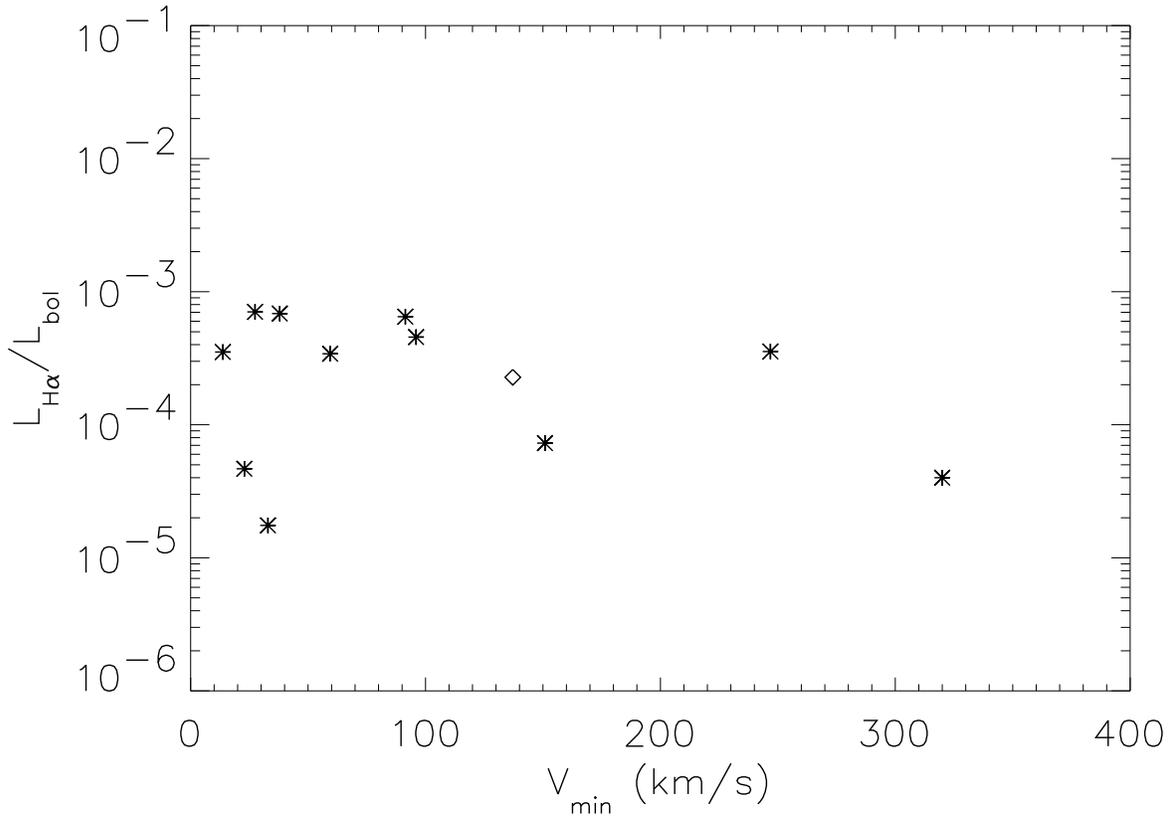}
\caption{Activity level vs V$_{min}$ (minimum $v \sin i$) for the 12 systems in Table 3,
based on the observed variation of the H$\alpha$ emission line.
SDSS112909.50$+$663704.4, shown as a diamond, has a 4.1 hour period (see
Figure 10).  Inclination
effects would move points to the right on this plot, i.e. to
higher true orbital velocities at the same activity level.}
\end{figure}

\begin{figure}
\plotone{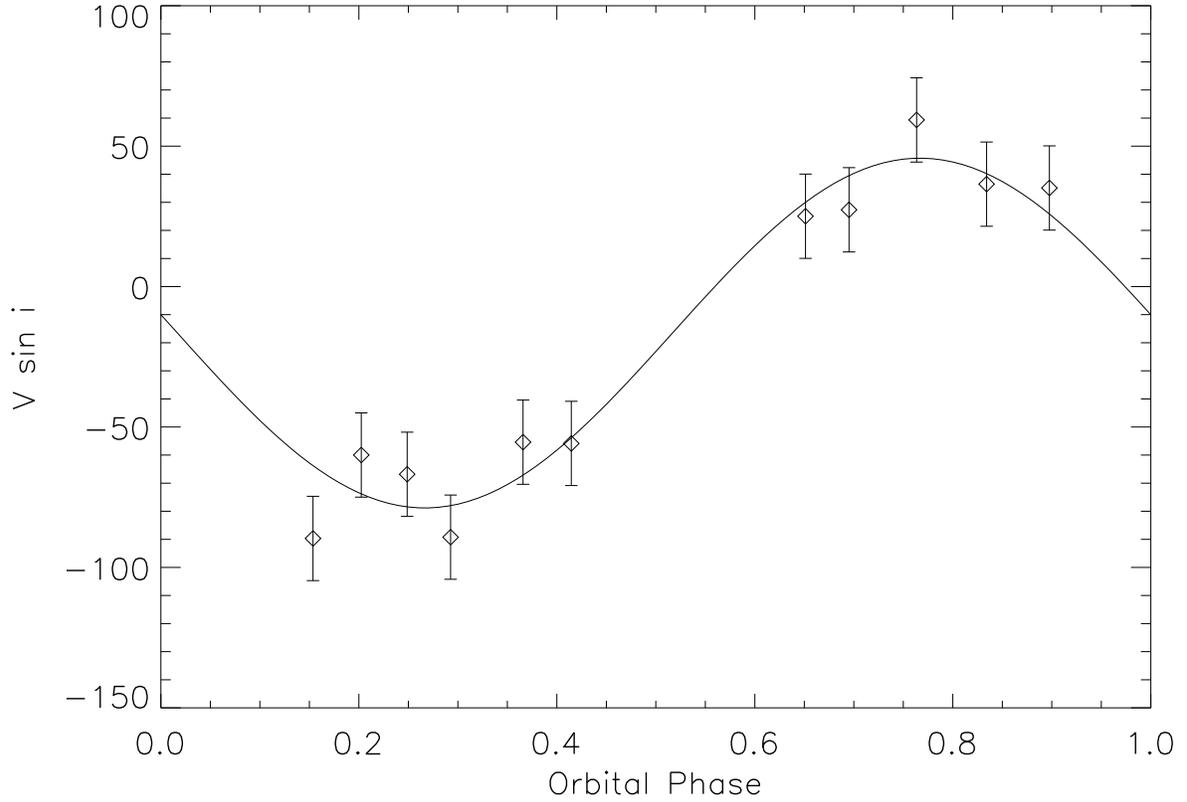}
\caption{Velocity curve for SDSS112909.50$+$663704.4, showing a 4.1 hour
period.  The velocities were calculated using the H$\alpha$ emission line.
The best-fit parameters are: P = 4.1 hrs $\pm$ 0.1 hrs, K = 62.3 km~s$^{-1}$ $\pm$
4.7 km~s$^{-1}$, $\gamma$ = -16.6 km~s$^{-1}$ $\pm$ 0.4 km~s$^{-1}$, and
$\phi_{0}$ = 0.017 $\pm$ 0.019.  Assuming a white dwarf mass of 0.6
M$_{\odot}$ and an M4.5 mass of 0.2 M$_{\odot}$ (Reid \& Hawley 2000), this
implies a separation of $\sim$ 10$^{11}$ cm ($\approx$ 1.4 R$_{\odot}$).
This places the system slightly outside the expected region for mass transfer.}
\end{figure}

\scriptsize
\begin{deluxetable}{lccrrrrrrr}
\tablewidth{0pt}
\tablecaption{WD/M Binary Parameters}
\tablehead{
\colhead{SDSS J} &  \colhead{{\it g}} & 
\colhead{{\it u-g}} &
\colhead{{\it g-r}} &  
\colhead{{\it r-i}} &
\colhead{{\it i-z}} & 
\colhead{T(WD)\tablenotemark{a}}  & 
\colhead{M sp type\tablenotemark{b}} &
\colhead{d(pc)\tablenotemark{c}} &
\colhead{EW[H$\alpha$]}}
\startdata

001726.64$-$002451.2  & 19.34 & 0.59 & 0.27 & 0.82 & 0.59 &  12500  &  5  &  87 	 &  4.2  \nl 
001733.59$+$004030.4\tablenotemark{g}  & 19.63 & 0.69 & 0.43 & 1.16 & 0.80 &  8500  &  5  &  57 	 &  0.0  \nl 
001749.25$-$000955.4\tablenotemark{df}  & 16.88 & -0.34 & -0.15 & 0.30 & 0.28 &  40000  &  2  &  90 	 &  8.6  \nl 
002750.00$-$001023.4  & 18.36 & 0.15 & 0.33 & 0.80 & 0.59 &  44000  &  4  &  193 	 &  0.0  \nl 
005208.42$-$005134.6\tablenotemark{g}  & 18.27 & 0.78 & 0.60 & 1.09 & 0.66 &  10000  &  5  &  42 	 &  0.0  \nl 
005457.61$-$002517.1\tablenotemark{g}  & 18.43 & 0.38 & -0.36 & 0.31 & 0.52 &  17000  &  7  &  86 	 &  3.1  \nl 
012259.53$+$154253.8  & 18.96 & 0.36 & -0.04 & 0.41 & 0.56 &  12500  &  5.5  &  75 	 &  0.0  \nl 
012527.14$+$010255.4  & 19.82 & 0.20 & 0.04 & 0.59 & 0.59 &  24000  &  3  &  191 	 &  3.0  \nl 
014152.51$-$005241.7\tablenotemark{g}  & 19.76 & 0.78 & 0.54 & 0.95 & 0.61 &  8000  &  5  &  68 	 &  23.0  \nl 
020351.29$+$004025.1\tablenotemark{dg}  & 19.35 & 0.93 & 0.68 & 1.09 & 0.60 &  9500  &  3  &  80 	 &  0.0  \nl 
020806.39$+$001834.0\tablenotemark{g}  & 17.25 & 0.67 & -0.37 & 0.67 & 1.02 &  10000  &  5  &  20 	 &  8.2  \nl 
021239.46$+$001856.9  & 19.41 & 0.53 & 0.36 & 1.02 & 0.64 &  17000  &  4  &  109 	 &  5.2  \nl 
021411.35$+$011533.3  & 19.88 & 0.65 & 0.81 & 0.85 & 0.46 &  9000  &  2.5  &  63 	 &  0.3  \nl 
021534.78$+$141846.7  & 19.55 & 0.01 & 0.15 & 0.39 & 0.36 &  24000  &  3  &  183 	 &  0.0  \nl 
022835.93$-$074032.4  & 18.75 & 0.36 & -0.12 & 0.37 & 0.56 &  15000  &  5.5  &  83 	 &  0.0  \nl 
023515.12$-$000737.0\tablenotemark{fg} & 16.71 & 0.24 & 0.51 & 0.84 & 0.54 &  9500  &  5  &  19 	 &  2.8  \nl 
024642.55$+$004137.2\tablenotemark{g}  & 19.15 & 0.84 & 0.74 & 1.14 & 0.63 &  9500  &  4  &  56 	 &  8.8  \nl 
025610.61$-$073024.6  & 16.83 & 0.41 & 0.00 & 0.58 & 0.61 &  14000  &  5.5  &  33 	 &  11.0  \nl 
025622.04$-$065036.2\tablenotemark{g}  & 19.45 & 0.42 & 0.08 & 0.59 & 0.52 &  13000  &  4.5  &  94 	 &  0.0  \nl 
025817.87$+$010946.0\tablenotemark{dfg} & 18.15 & -0.12 & 0.06 & 0.68 & 0.65 &  28000  &  3.5  &  110 	 &  2.8  \nl 
030302.44$-$062829.9\tablenotemark{g}  & 19.77 & 0.44 & 0.13 & 0.61 & 0.63 &  15000  &  4.5  &  134 	 &  1.4  \nl 
030544.41$-$074941.2  & 18.31 & 0.36 & 0.19 & 0.83 & 0.62 &  15000  &  5  &  69 	 &  6.4  \nl 
030607.19$-$003114.4\tablenotemark{fg}  & 16.39 & 0.29 & 0.20 & 0.82 & 0.60 &  15000  &  5  &  32 	 &  9.3  \nl 
030859.87$-$002735.8  & 19.64 & 0.75 & 0.73 & 1.06 & 0.64 &  9500  &  4.5  &  75 	 &  8.7  \nl 
032136.55$-$001630.5\tablenotemark{g}  & 17.62 & -0.16 & -0.12 & 0.61 & 0.55 &  22000  &  4.5  &  64 	 &  6.1  \nl 
032422.38$-$000516.1\tablenotemark{g}  & 19.33 & 0.01 & 0.08 & 0.66 & 0.62 &  18000  &  4.5  &  126 	 &  10.5  \nl 
032428.78$-$004613.8\tablenotemark{g}  & 19.30 & 0.09 & 0.38 & 0.43 & 0.03 &  32000  &  4  &  233 	 &  1.5  \nl 
032758.16$-$002215.5\tablenotemark{g}  & 19.52 & 0.96 & 0.88 & 1.08 & 0.64 &  10500  &  4  &  82 	 &  0.9  \nl 
032842.92$+$001749.8\tablenotemark{g}  & 18.05 & 0.94 & 0.92 & 0.98 & 0.58 &  11000  &  2.5  &  46 	 &  0.0  \nl 
033548.59$+$003832.3\tablenotemark{dg}  & 18.02 & 0.23 & 0.10 & 0.82 & 0.65 &  17000  &  4.5  &  62 	 &  0.0  \nl 
034831.09$-$061400.5\tablenotemark{g} & 19.63 & 0.33 & 0.01 & 0.65 & 0.60 &  15000  &  4.5  &  114 	 &  0.0  \nl 
042328.12$-$003801.5  & 19.56 & 0.05 & 0.33 & 0.95 & 0.66 &  8000  &  3.5  &  43 	 &  27.0  \nl 
074011.18$+$385909.1  & 19.06 & 0.32 & -0.16 & 0.07 & 0.33 &  12000  &  6  &  80 	 &  0.0  \nl 
074301.93$+$410655.3\tablenotemark{f}  & 17.32 & 0.14 & -0.14 & 0.47 & 0.58 &  16000  &  5.5  &  43 	 &  3.1  \nl 
074730.57$+$430403.6\tablenotemark{f}  & 18.75 & 0.50 & 0.41 & 0.92 & 0.75 &  9500  &  3  &  51 	 &  5.4  \nl 
075009.90$+$381641.6  & 18.23 & 0.18 & -0.09 & 0.57 & 0.56 &  11500  &  5  &  47 	 &  4.0  \nl 
075223.15$+$433212.2  & 19.12 & 0.34 & 0.57 & 0.70 & 0.51 &  24000  &  4.5  &  158 	 &  5.3  \nl 
083808.00$+$530254.4  & 19.39 & -0.07 & 0.43 & 0.61 & 0.54 &  24000  &  2.5  &  40 	 &  4.4  \nl 
084833.60$+$005840.0  & 18.78 & 0.30 & 0.26 & 0.91 & 0.67 &  12000  &  5  &  70 	 &  2.5  \nl 
090925.42$+$533700.7  & 17.37 & 0.11 & -0.01 & 0.57 & 0.56 &  17000  &  4  &  51 	 &  2.7  \nl 
091849.56$+$545601.8  & 18.25 & 0.20 & 0.66 & 0.60 & 0.41 &  8000  &  1  &  26 	 &  0.4  \nl 
092451.63$-$001736.5  & 18.57 & 0.24 & 0.36 & 0.79 & 0.58 &  9500  &  5  &  42 	 &  8.8  \nl 
092452.40$+$002449.0  & 18.04 & 0.19 & -0.20 & 0.35 & 0.50 &  18000  &  5  &  70 	 &  2.0  \nl 
093918.47$+$572901.2  & 18.33 & 0.38 & 0.12 & 0.71 & 0.60 &  11000  &  4  &  44 	 &  0.0  \nl 
095108.74$+$025507.5  & 18.87 & 0.45 & 0.00 & 0.49 & 0.58 &  12000  &  5  &  72 	 &  0.0  \nl 
095706.00$+$615254.2  & 18.82 & -0.26 & -0.02 & 0.61 & 0.48 &  30000  &  2  &  156 	 &  0.0  \nl 
100609.18$+$004417.1\tablenotemark{dg}  & 17.85 & 0.44 & 0.12 & 0.33 & 0.65 &  8000  &  7  &  21 	 &  0.0  \nl 
100953.70$-$002853.5\tablenotemark{g}  & 19.42 & 0.21 & -0.06 & 0.60 & 0.64 &  19000  &  4.5  &  130 	 &  1.6  \nl 
101647.21$-$010907.2  & 17.40 & 0.24 & -0.28 & 0.47 & 0.68 &  17000  &  5.5  &  43 	 &  8.3  \nl 
103743.61$+$655448.0  & 18.36 & 0.38 & -0.07 & 0.36 & 0.52 &  15000  &  6  &  63 	 &  14.8  \nl 
104517.78$-$001833.7\tablenotemark{g}  & 18.33 & 0.14 & -0.17 & 0.53 & 0.65 &  15000  &  5  &  65 	 &  0.7  \nl 
105853.59$+$005608.6\tablenotemark{g}  & 18.93 & 0.11 & 0.33 & 0.37 & 0.28 &  19000  &  0  &  109 	 &  0.0  \nl 
112307.36$+$003216.1  & 19.42 & 0.19 & -0.13 & 0.62 & 0.58 &  15000  &  6  &  112 	 &  3.9  \nl 
112357.73$-$011027.5  & 18.30 & 0.10 & -0.28 & 0.13 & 0.50 &  15000  &  6.5  &  68 	 &  0.0  \nl 
112623.88$+$010856.9\tablenotemark{g}  & 18.71 & 0.42 & 0.49 & 0.87 & 0.54 &  9500  &  3  &  46 	 &  3.5  \nl 
112748.27$-$002859.9\tablenotemark{g}  & 18.34 & 0.37 & -0.21 & 0.39 & 0.60 &  17000  &  5.5  &  78 	 &  0.8  \nl 
112909.50$+$663704.4\tablenotemark{f}  & 17.70 & 0.26 & -0.05 & 0.54 & 0.66 &  17000  &  4.5  &  62 	 &  12.4  \nl 
113457.72$+$655408.7  & 18.24 & 0.02 & 0.07 & 0.58 & 0.54 &  12500  &  6  &  59 	 &  4.5  \nl 
113722.25$+$014858.6\tablenotemark{h}  & 18.74 & -0.08 & 0.15 & 0.38 & 0.59 &  42000  &  2.5  &  182 	 &  100.0  \nl 
113800.35$-$001144.5\tablenotemark{dg}  & 18.82 & 0.30 & -0.04 & 0.71 & 0.64 &  26000  &  5  &  135 	 &  0.0  \nl 
114312.57$+$000926.5\tablenotemark{dg}  & 18.13 & 0.23 & -0.11 & 0.51 & 0.65 &  16000  &  4.5  &  62 	 &  3.2  \nl 
115156.94$-$000725.5\tablenotemark{g}  & 18.11 & 0.46 & -0.03 & 0.33 & 0.51 &  10500  &  6.5  &  41 	 &  11.0  \nl 
120507.97$+$031234.4  & 18.47 & 0.38 & -0.05 & 0.46 & 0.65 &  12000  &  5  &  57 	 &  0.0  \nl 
121033.77$-$004644.4\tablenotemark{dg}  & 18.92 & 0.24 & 0.53 & 1.11 & 0.67 &  8000  &  4  &  40 	 &  0.0  \nl 
122339.62$-$005631.2\tablenotemark{g}  & 17.98 & 0.51 & -0.16 & 0.15 & 0.34 &  13500  &  5.5  &  47 	 &  0.6  \nl 
123922.34$+$005548.8\tablenotemark{dg}  & 19.24 & 0.27 & 0.03 & 0.83 & 0.67 &  16000  &  4.5  &  110 	 &  5.7  \nl 
125207.17$+$012156.8  & 18.78 & -0.03 & 0.28 & 0.58 & 0.41 &  9000  &  3  &  41 	 &  0.0  \nl 
125250.03$-$020608.2  & 19.21 & 0.12 & 0.23 & 0.61 & 0.53 &  22000  &  1.5  &  158 	 &  0.0  \nl 
132312.65$-$025456.1  & 18.02 & 0.33 & 0.59 & 1.09 & 0.69 &  8000  &  3.5  &  29 	 &  0.0  \nl 
133953.89$+$015614.9  & 19.34 & 0.31 & 0.13 & 0.59 & 0.54 &  10500  &  3  &  71 	 &  6.9  \nl 
140723.04$+$003841.8\tablenotemark{d}  & 17.66 & 0.04 & -0.09 & 0.74 & 0.60 &  24000  &  4  &  76 	 &  3.7  \nl 
141220.70$+$654123.3  & 19.17 & 0.35 & -0.11 & 0.43 & 0.48 &  13000  &  5  &  84 	 &  2.0  \nl 
141731.25$+$015059.0  & 18.71 & 0.37 & 0.02 & 0.60 & 0.54 &  13500  &  4.5  &  69 	 &  9.8  \nl 
143947.62$-$010606.9\tablenotemark{f}  & 16.55 & -0.36 & -0.27 & -0.01 & 0.11 &  40000  &  0  &  73 	 &  10.7  \nl 
144258.48$+$001031.5\tablenotemark{f}  & 18.29 & -0.01 & 0.03 & 0.78 & 0.72 &  32000  &  4.5  &  132 	 &  4.5  \nl 
145300.99$+$005557.1\tablenotemark{d}  & 19.16 & 0.41 & 0.04 & 0.61 & 0.57 &  13000  &  5  &  80 	 &  2.0  \nl 
145527.72$+$611504.8  & 19.24 & 0.03 & -0.01 & 0.66 & 0.54 &  12500  &  5  &  94 	 &  0.0  \nl 
150118.41$+$042232.3  & 19.58 & -0.05 & -0.01 & 0.48 & 0.51 &  16000  &  5.5  &  133 	 &  4.6  \nl 
150231.66$+$011045.9\tablenotemark{e}  & 18.48 & 0.41 & -0.03 & 0.43 & 0.56 &  13000  &  4  &  60 	 &  0.0  \nl 
150648.12$+$002716.4  & 19.05 & 0.24 & 0.53 & 0.73 & 0.49 &  8500  &  3  &  41 	 &  9.0  \nl 
151058.76$+$004112.7  & 19.39 & 0.45 & -0.05 & 0.36 & 0.64 &  18000  &  6  &  125 	 &  16.0  \nl 
152717.25$+$002413.4\tablenotemark{g}  & 19.60 & 0.55 & -0.04 & 0.63 & 0.61 &  10000  &  6.5  &  61 	 &  0.0  \nl 
152933.26$+$002031.2\tablenotemark{dg}  & 18.18 & 0.46 & -0.15 & 0.34 & 0.53 &  16000  &  5.5  &  62 	 &  15.3  \nl 
154219.45$+$032311.0  & 19.46 & 0.34 & 0.34 & 0.98 & 0.59 &  10500  &  3.5  &  76 	 &  0.0  \nl 
155554.25$+$545417.4  & 18.86 & -0.09 & 0.03 & 0.54 & 0.42 &  20000  &  2.5  &  116 	 &  0.0  \nl 
161033.87$+$491536.7  & 18.77 & 0.04 & -0.05 & 0.48 & 0.46 &  19000  &  3.5  &  99 	 &  0.0  \nl 
161145.89$+$010327.8  & 18.76 & 0.77 & 0.09 & 0.59 & 0.67 &  20000  &  5  &  96 	 &  9.1  \nl 
164615.60$+$422349.3  & 18.43 & 0.31 & -0.13 & 0.59 & 0.75 &  15000  &  6.5  &  74 	 &  16.6  \nl 
165005.16$+$390426.4\tablenotemark{g}  & 18.73 & -0.06 & -0.15 & 0.55 & 0.55 &  15000  &  5.5  &  91 	 &  9.0  \nl 
170459.71$+$384433.0  & 18.34 & 0.18 & -0.04 & 0.47 & 0.49 &  19000  &  4  &  77 	 &  0.0  \nl 
171145.43$+$555444.5\tablenotemark{dg}  & 18.24 & -0.32 & -0.34 & 0.32 & 0.51 &  32000  &  4.5  &  130 	 &  1.5  \nl 
171301.85$+$625135.9\tablenotemark{g}  & 19.31 & 0.31 & 0.35 & 0.80 & 0.54 &  8000  &  3  &  38 	 &  7.3  \nl 
171611.13$+$544346.5\tablenotemark{dg}  & 19.05 & 0.19 & 0.20 & 0.75 & 0.49 &  18000  &  2.5  &  120 	 &  2.8  \nl 
171810.16$+$610114.0\tablenotemark{g}  & 18.00 & 0.15 & -0.24 & 0.43 & 0.64 &  17000  &  6  &  65 	 &  12.9  \nl 
171955.23$+$625106.8\tablenotemark{g}  & 19.34 & 0.24 & -0.10 & 0.68 & 0.82 &  26000  &  4.5  &  65 	 &  3.0  \nl 
172043.87$+$560109.3\tablenotemark{g}  & 18.76 & 0.82 & 1.03 & 0.82 & 0.59 &  10000  &  3  &  51 	 &  3.4  \nl 
172433.70$+$623410.0\tablenotemark{g}  & 17.43 & -0.07 & -0.44 & 0.42 & 0.41 &  9500  &  2.5  &  54 	 &  4.1  \nl 
172441.80$+$593103.3\tablenotemark{fg}  & 17.76 & 0.03 & -0.15 & 0.59 & 0.56 &  15000  &  5  &  59 	 &  10.2  \nl 
172601.55$+$560527.1\tablenotemark{dg}  & 19.20 & 0.22 & 0.30 & 0.76 & 0.49 &  16000  &  3.5  &  101 	 &  4.4  \nl 
172715.57$+$542938.1\tablenotemark{g}  & 19.37 & 0.47 & 0.47 & 0.87 & 0.55 &  11500  &  3.5  &  86 	 &  4.5  \nl 
172831.84$+$620426.5\tablenotemark{g}  & 18.80 & 0.34 & 0.59 & 0.89 & 0.60 &  16000  &  5  &  72 	 &  14.0  \nl 
173101.49$+$623316.0\tablenotemark{g}  & 18.18 & 0.31 & -0.12 & 0.57 & 0.65 &  17000  &  5.5  &  118 	 &  0.0  \nl 
214720.96$+$130227.4  & 19.62 & 0.06 & 0.03 & 0.52 & 0.52 &  17000  &  4  &  147 	 &  0.0  \nl 
221714.49$+$004601.3  & 19.42 & 0.36 & 0.10 & 0.60 & 0.71 &  17000  &  5.5  &  113 	 &  0.0  \nl 
231221.60$+$010127.0  & 18.66 & 0.19 & 0.61 & 1.02 & 0.66 &  20000  &  3.5  &  128 	 &  0.0  \nl 
231230.79$+$005321.7\tablenotemark{f}  & 17.23 & 0.13 & 0.53 & 0.71 & 0.51 &  22000  &  3  &  64 	 &  8.8  \nl 
233919.65$-$000233.8  & 19.14 & 0.91 & 0.74 & 0.92 & 0.58 &  9000  &  3.5  &  53 	 &  0.0  \nl 
234459.62$+$002750.0  & 19.42 & 0.40 & -0.13 & 1.82 & 0.94 &  8000  &  5.5  &  52 	 &  5.0  \nl 
234649.94$-$002032.2  & 18.15 & 0.27 & 0.30 & 0.64 & 0.36 &  9500  &  2.5  &  30 	 &  3.0  \nl 

\enddata
\tablenotetext{a}{White Dwarf Primary Temperature (K): Errors are $\pm$2000K
for T $\leq$ 20,000K, $\pm$4000K for T $>$ 20,000K}
\tablenotetext{b}{Secondary M Spectral Type: Errors are $\pm$1-2 subclasses}
\tablenotetext{c}{Distance errors are $\pm$ 20 \%}
\tablenotetext{d}{These objects have been observed twice by SDSS}
\tablenotetext{e}{This object has been observed three times by SDSS}
\tablenotetext{f}{These objects have followup APO observations.  See Table 3.}
\tablenotetext{g}{These objects are in the SDSS Early Data Release (EDR).  See
Stoughton et al 2002.}
\tablenotetext{h}{This object is a CV discovered by our photometric search pipeline.}
\end{deluxetable}

\clearpage

\scriptsize
\begin{deluxetable}{lrrcl}
%\tablewidth{0pt}
\tablecaption{M Dwarf Secondary Magnetic Activity Data}
\tablehead{
\colhead{M spectral type}  & \colhead{Objects\tablenotemark{a}} &
\colhead{Active Objects\tablenotemark{b}} &
\colhead{Log[Mean(L$_{H\alpha}$/L$_{bol}$)]\tablenotemark{c}} &
\colhead{Error(dex)\tablenotemark{d}}}
\startdata

0 & 2 & 0 & -- & -- \nl
0.5 & 0 & 0 & -- & -- \nl
1 & 1 & 1 & -4.29 & 0.76 \nl
1.5 & 1 & 0 & -- & -- \nl
2 & 2 & 0 & -- & -- \nl
2.5 & 8 & 6 & -3.02 & 0.08 \nl
3 & 11 & 9 & -2.61 & 0.21 \nl
3.5 & 9 & 7 & -2.76 & 0.10 \nl
4 & 12 & 8 & -2.47 & 0.44 \nl
4.5 & 16 & 9 & -3.33 & 0.33 \nl
5 & 22 & 10 & -3.49 & 0.36 \nl
5.5 & 13 & 9 & -3.77 & 0.62 \nl
6 & 6 & 4 & -4.98 & 2.08 \nl
6.5 & 4 & 1 & -4.11 & 1.74 \nl
7 & 2 & 0 & -- & -- \nl

\enddata

\tablenotetext{a}{Number of Objects in spectral type bin.}
\tablenotetext{b}{Number of Objects in spectral type bin which are
  magnetically active.}
\tablenotetext{c}{Mean activity level of the active objects in each spectral
  type bin.} 
\tablenotetext{d}{Error in the logarithm of the mean L$_{H\alpha}$/L$_{bol}$
  for each spectral type bin.}

\end{deluxetable}

\clearpage

\scriptsize
\begin{deluxetable}{lrrrrl}
%\tablewidth{0pt}
\tablecaption{APO Followup Data\tablenotemark{a}}
\tablehead{
\colhead{SDSS J}  & \colhead{Nspec\tablenotemark{b}} &
\colhead{Hours\tablenotemark{c}} &
\colhead{Dates\tablenotemark{d}} &
\colhead{V$_{min}$\tablenotemark{e}} & \colhead{Comments}}
\startdata

001749.25$-$000955.4 & 3 & $-$ & E,F & 27 & very blue SED \nl 
023515.12$-$000737.0 & 3 & $-$ & E,F & 32 & \nl
025817.87$+$010946.0 & 3 & $-$ & E,F & 14 & \nl
030607.19$-$003114.4 & 12 & 2 & A,E,F & 320 & \nl
074301.93$+$410655.3 & 3 & $-$ & G & 23 & \nl
074730.57$+$430403.7 & 19 & 4 & B,E,G,H & 250 & \nl
075223.15$+$433212.2 & 5 & $-$ & E,G,H & 96 & \nl
112909.50$+$663704.4 & 17 & 3 & B,C,G & 137 & 4.1 hour period\nl
143947.62$-$010606.9 & 25 & 5.5 & C & 91 & \nl
144258.48$-$001031.5 & 2 & $-$ & B,D & 150 & \nl
172441.80$+$593103.3 & 9 & 2.5 & C & 59 & \nl
231230.79$+$005321.7 & 3 & $-$ & D,F & 27 & \nl

\enddata

\tablenotetext{a}{Finding charts were obtained from the STScI Digitized Sky
  Survey at  http://archive.stsci.edu/cgi-bin/dss\_form}
\tablenotetext{b}{Number of APO spectra taken}
\tablenotetext{c}{Hours of continuous observation on a single night}
\tablenotetext{d}{Dates of APO spectra taken: A=9/13/2000, B=4/15/2001,
C=5/4/2001, D=6/17/2001, E=10/15/2001, F=12/7/2001, G=12/18/2001, H=12/19/2001}
\tablenotetext{e}{Minimum radial velocity variation in km~s$^{-1}$, derived from the
maximum wavelength variation in the H$\alpha$ emission line.}

\end{deluxetable}

\clearpage

\acknowledgments

\end{document}